\documentclass[a4paper,11pt]{article}
\usepackage{pos}
\usepackage{lineno}
\title{Measurement of semi-inclusive jet fragmentation functions in Au+Au collisions at $\sqrt{s_{\mathrm{NN}}} = 200$ GeV in STAR}

\author*[a]{Saehanseul Oh (for the STAR Collaboration)}

\affiliation[a]{Lawrence Berkeley National Laboratory,\\
1 Cyclotron Rd, Berkeley, CA}

\emailAdd{saehanseul.oh@lbl.gov}

\abstract{
Jet quenching in relativistic heavy ion collisions can have multiple phenomenological consequences: jet energy loss, modification of jet substructure, and induced acoplanarity.
In these proceedings, we report a measurement of the jet fragmentation function, which is one of the jet substructure observables, in peripheral Au+Au collisions at $\sqrt{s_{\mathrm{NN}}}=200$ GeV by the STAR experiment at RHIC. 
In particular, we use a semi-inclusive population of jets recoiling from a high transverse momentum trigger hadron. 
The fragmentation function is constructed from the fraction of the transverse momentum of charged particles projected onto the jet axis over the transverse momentum of the jet. 
In a previous STAR publication of the semi-inclusive charged-jet spectra, the Mixed-Event technique was used along with the semi-inclusive approach to remove the uncorrelated background contributions, which enables the measurement of jet distributions at low transverse momentum and large jet radius ($R$) in heavy ion collisions. 
In this analysis we extend this approach to the measurement of jet fragmentation functions. 
The reported fragmentation functions are corrected for uncorrelated background and instrumental effects via unfolding.
The results are compared to those in PYTHIA simulations for $pp$ collisions. 

\FullConference{%
  HardProbes2020\\
  1-6 June 2020\\
  Austin, Texas}
}

\begin{document}
\maketitle

\section{Introduction}
\label{sec:intro}
Jet quenching refers to modifications of a jet shower in Quark-Gluon Plasma (QGP) generated in relativistic heavy-ion collisions, and is one of the key observables used to study the properties of QGP~\cite{Busza:2018rrf}. 
The modification of jet substructure due to jet quenching has been investigated at the LHC, with various observables, such as jet shapes, fragmentation functions, and shared momentum fraction. 
With increased luminosity from the recent RHIC runs and advanced techniques in handling background jets~\cite{Adamczyk:2016fqm, Adamczyk:2017yhe}, such measurements have become feasible at RHIC energies with the STAR detector. 
In these proceedings, we present the measurement of semi-inclusive jet fragmentation functions in peripheral Au+Au collisions at $\sqrt{s_{\mathrm{NN}}}=200$ GeV. 

Jet fragmentation functions are defined as, 
\begin{eqnarray}
D(p_{\mathrm{T,jet}}, z) = \frac{1}{N_{\mathrm{jet}}} \frac{\mathrm{d}N_{\mathrm{ch}}(p_{\mathrm{T,jet}}, z)}{\mathrm{d}z}\,,
\label{eq:FF}
\end{eqnarray}
where $z = p_{\mathrm{T,track}}\,\mathrm{cos}(\Delta R)/p_{\mathrm{T,jet}}$ and  $\Delta R = \sqrt{(\varphi_{\mathrm{jet}} - \varphi_{\mathrm{track}})^2 + (\eta_{\mathrm{jet}} - \eta_{\mathrm{track}})^2}$, and for tracks satisfying $\Delta R$ < 0.4. 
$p_{\mathrm{T}}$, $\eta$, and $\varphi$ represent the transverse momentum, pseudo-rapidity, and azimuthal angle, respectively. 
This corresponds to the distribution of constituent charged particle's longitudinal momentum fraction with respect to the jet momentum normalized per jet.
In-medium modifications of fragmentation functions have been previously reported by LHC collaborations for inclusive jet populations~\cite{Aaboud:2017bzv, Chatrchyan:2012gw}. 
In the semi-inclusive approach, the fragmentation functions are reported for the population of recoil jets with respect to a high momentum trigger particle.

\section{Measurement}
\label{sec:Measurement}
For this analysis, Au+Au collisions at $\sqrt{s_{\mathrm{NN}}}=$ 200 GeV, collected in 2014 by the STAR experiment, are used~\cite{Ackermann:2002ad}.
Charged tracks with $0.2 < p_{\mathrm{T,track}} < 30.0$ GeV/$c$ and $|\eta_{\mathrm{track}}| < 1.0$ are reconstructed with the Time Projection Chamber (TPC), and events containing a track with $p_{\mathrm{T,track}}$ higher than 30.0 GeV/$c$ are excluded in the measurement. 
Jets are reconstructed with these charged tracks using the anti-$k_{\mathrm{T}}$ sequential jet clustering algorithm from the FastJet package~\cite{Cacciari:2011ma} with a resolution parameter $R =$ 0.4. 
A semi-inclusive approach is used following procedures described in~\cite{Adamczyk:2017yhe}, and jets in the recoil region ($3\pi/4 < \vert \varphi_{\mathrm{jet}} - \varphi_{\mathrm{trig}} \vert < 5\pi/4$) of a high energy Barrel Electromagnetic Calorimeter (BEMC) trigger tower ($9.0 < E_{\mathrm{T}} < 30.0$ GeV) are considered. 
While signal jets, i.e.~correlated jets with respect to the trigger tower, are measured from events triggered by a high energy BEMC tower, the uncorrelated background contributions are estimated using minimum-bias events via a mixed-event technique. 
Mixed events are constructed for distinct classes of events based on their reconstructed charged-particle multiplicity, primary vertex position along the beam direction, and the second-order event-plane angle.

For the measurement of jet fragmentation functions (Eq.~\ref{eq:FF}), components that are uncorrelated with the trigger tower in $N_{\mathrm{jet}}$ and those in $\mathrm{d}N_{\mathrm{ch}}/\mathrm{d}z$ are removed independently. 
For $N_{\mathrm{jet}}$, the number of jets in each $p_{\mathrm{T,jet}}$ bin, the same subtraction procedures as \cite{Adamczyk:2017yhe} are applied. 
For $\mathrm{d}N_{\mathrm{ch}}/\mathrm{d}z$, contributions from 1) uncorrelated jets, and 2) uncorrelated particles in correlated jets are separately evaluated using mixed events. 
For 1), $\mathrm{d}N_{\mathrm{ch}}^{\mathrm{ME}}/\mathrm{d}z$ is measured with mixed events, and scaled according to the uncorrelated jet fraction in each $p_{\mathrm{T,jet}}$ bin. 
For 2), correlated jets are placed into mixed events, and combined with mixed-event tracks that satisfy $\Delta R$ < 0.4. 
These contributions from 1) and 2) are subtracted from $\mathrm{d}N_{\mathrm{ch}}/\mathrm{d}z$ from all recoil jets. 
After such subtractions, $N_{\mathrm{jet}}$ and $\mathrm{d}N_{\mathrm{ch}}/\mathrm{d}z$ are independently unfolded via 1-dimensional and 2-dimensional Bayesian unfolding~\cite{Adye:2011gm}, respectively, for uncorrelated background effects and instrumental effects in the fragmentation functions.

\section{Results}
\label{sec:Results}
\begin{figure}[h]
	\centering
	\includegraphics[width=0.56 \textwidth]{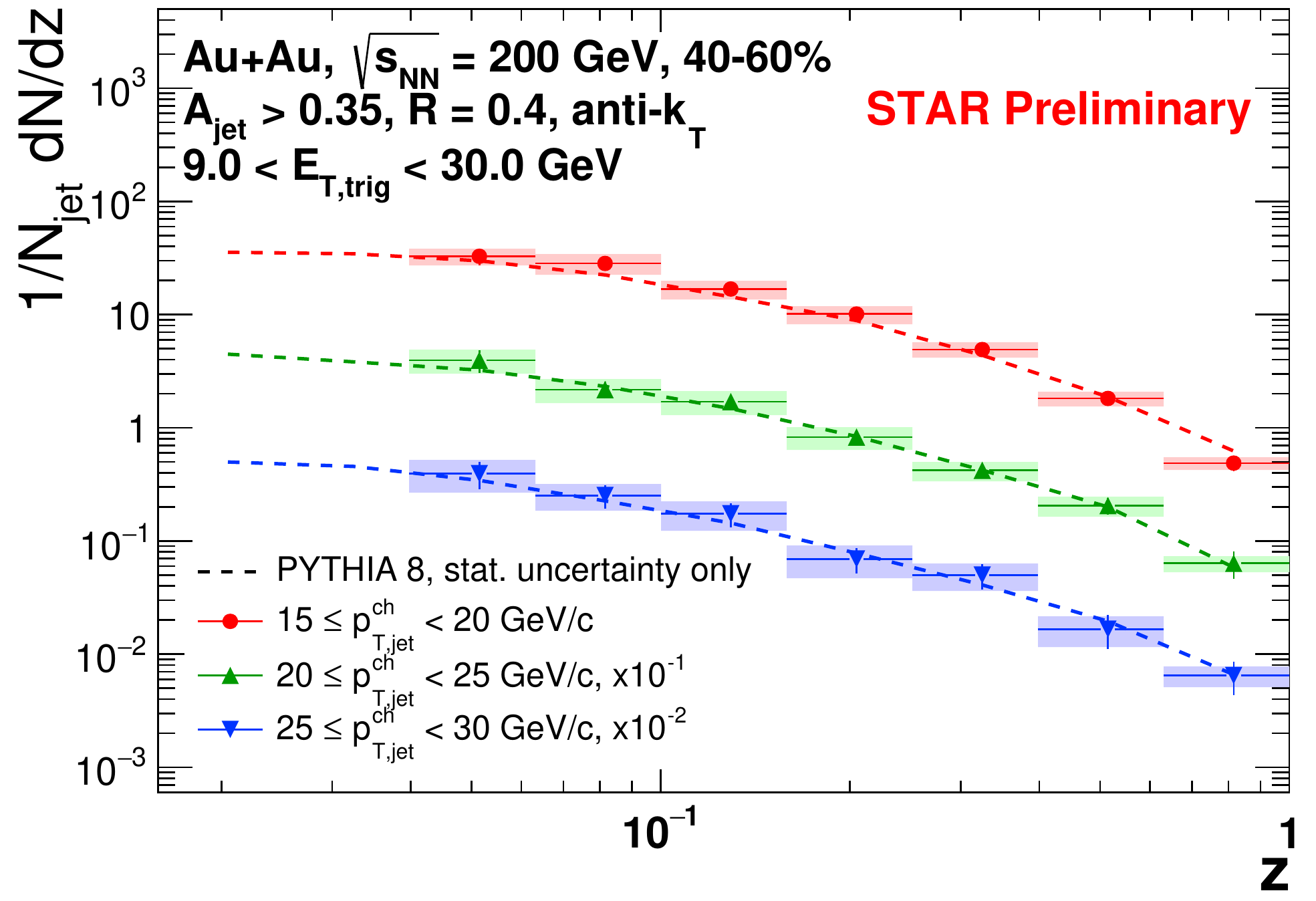}
	\caption{\label{fig:3} Semi-inclusive jet fragmentation functions measured in 40--60\% Au+Au collisions for three $p_{\mathrm{T,jet}}$ ranges (closed markers), compared to those calculated by PYTHIA 8 for $pp$ collisions (dashed lines). Measured distributions are corrected for detector effects and uncorrelated background effects.}
\end{figure}

\begin{figure}[h]
	\centering
	\includegraphics[width=0.84\textwidth]{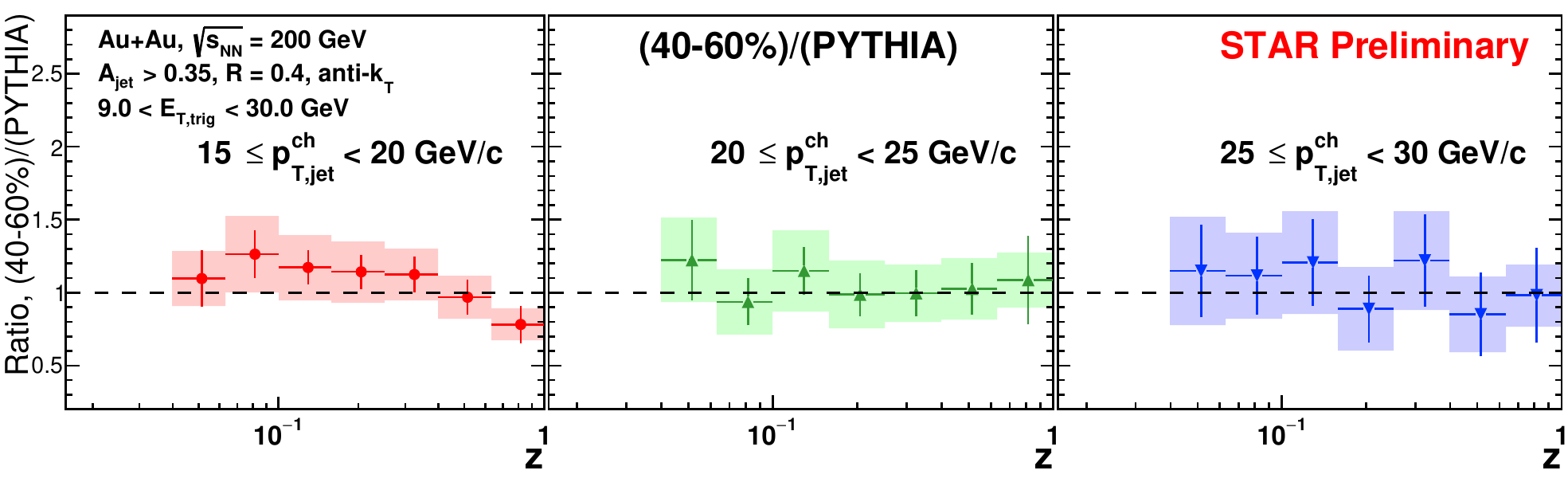}
	\caption{\label{fig:4} Ratios of jet fragmentation functions measured in 40--60\% Au+Au collisions at $\sqrt{s_{\mathrm{NN}}} = 200$ GeV to those simulated by PYTHIA 8 for $pp$ collisions for three $p_{\mathrm{T,jet}}^{\mathrm{ch}}$ ranges.}
\end{figure}

Figure~\ref{fig:3} shows jet fragmentation functions for 40-60\% Au+Au collisions and three $p_{\mathrm{T,jet}}$ ranges, along with PYTHIA 8 (Monash 2013 tune~\cite{Skands:2014pea}) predictions for $pp$ collisions \cite{Sjostrand:2014zea}. 
Figure \ref{fig:4} shows the ratios between 40-60\% Au+Au collisions and PYTHIA 8 $pp$ estimations.
The observed ratios are consistent with unity within uncertainties over the measured $z$ and $p_{\mathrm{T,jet}}$ ranges, indicating no significant modification of jet fragmentation functions in 40-60\% Au+Au collisions at $\sqrt{s_{\mathrm{NN}}} = 200$ GeV. 
These results can be connected to various scenarios, such as no significant jet-medium interactions in 40-60\% heavy-ion collisions at $\sqrt{s_{\mathrm{NN}}}=200$ GeV collisions, or the possibility that PYTHIA 8 results in Figs.~\ref{fig:3} and \ref{fig:4} may not accurately represent the $pp$ events at $\sqrt{s} = 200$ GeV, as Monash 2013 tune is based on LHC data. 
Further studies are needed to validate the above physics case, as well as the $pp$ reference.

\section{Outlook}
\label{sec:Outlook}
In these proceedings, semi-inclusive jet fragmentation functions are reported for 40-60\% Au+Au collisions. 
The results in this centrality are observed to be consistent with those of PYTHIA 8 predictions for $pp$ collisions. 
In the future, these measurements will be extended to the most central Au+Au collisions, and compared to the corresponding measurements in $pp$ collisions. 
They will elucidate medium-induced modification of jet substructure at RHIC energies. 

\section*{Acknowledgement}
This work was supported by the Director, Office of Science, Office of Basic Energy Sciences, of the U.S.~Department of Energy under Contract No.~DE-AC02-05CH11231.


\begin{thebibliography}{99}
\bibitem{Busza:2018rrf} W.~Busza, K.~Rajagopal and W.~van der Schee, ``Heavy Ion Collisions: The Big Picture, and the Big Questions'', Ann. Rev. Nucl. Part. Sci. \textbf{68}, 339-376 (2018).
\bibitem{Adamczyk:2016fqm} L.~Adamczyk \textit{et al.}, ``Dijet imbalance measurements in $Au+Au$ and $pp$ collisions at $\sqrt{s_{NN}} = 200$  GeV at STAR'', Phys. Rev. Lett. \textbf{119}, no.6, 062301 (2017).
\bibitem{Adamczyk:2017yhe} L.~Adamczyk \textit{et al.}, ``Measurements of jet quenching with semi-inclusive hadron+jet distributions in Au+Au collisions at $\sqrt{s_{NN}}$ = 200 GeV'', Phys. Rev. C \textbf{96}, no.2, 024905 (2017).
\bibitem{Aaboud:2017bzv} M.~Aaboud \textit{et al.}, ``Measurement of jet fragmentation in Pb+Pb and $pp$ collisions at $\sqrt{{s_\mathrm{NN}}} = 2.76$ TeV with the ATLAS detector at the LHC'', Eur. Phys. J. C \textbf{77}, no.6, 379 (2017).
\bibitem{Chatrchyan:2012gw} S.~Chatrchyan \textit{et al.}, ``Measurement of jet fragmentation into charged particles in $pp$ and PbPb collisions at $\sqrt{s_{NN}}=2.76$ TeV'', JHEP \textbf{10}, 087 (2012).
\bibitem{Ackermann:2002ad} K.~H.~Ackermann \textit{et al.}, ``STAR detector overview'', Nucl. Instrum. Meth. A \textbf{499}, 624-632 (2003).
\bibitem{Cacciari:2011ma} M. Cacciari, G. P. Salam, G. Soyez, ``FastJet user manual'', Eur. Phys. J. C72 (2012) 1896.
\bibitem{Adye:2011gm} T. Adye, Proceedings: PHYSTAT 2011 Workshop on Statistical Issues Related to Discovery Claims in Search Experiments and Unfolding, CERN,Geneva, Switzerland 17-20 January 2011, 313-318.
\bibitem{Sjostrand:2014zea} T.~Sjostrand, S.~Ask, J.~R.~Christiansen, R.~Corke, N.~Desai, P.~Ilten, S.~Mrenna, S.~Prestel, C.~O.~Rasmussen and P.~Z.~Skands, ``An introduction to PYTHIA 8.2'', Comput. Phys. Commun. \textbf{191}, 159-177 (2015).
\bibitem{Skands:2014pea} P.~Skands, S.~Carrazza and J.~Rojo, ``Tuning PYTHIA 8.1: the Monash 2013 Tune'', Eur. Phys. J. C \textbf{74}, no.8, 3024 (2014).



%

\end{thebibliography}
\end{document}